# Индекс подобия математических и других научных публикаций с уравнениями и формулами и проблема идентификации самоплагиата

© А.Д. Полянин[1,2], И.К. Шингарева[3]

[1]Институт проблем механики им. А.Ю. Ишлинского РАН, Москва, 119526, Россия
[2]Московский государственный технический университет им. Н.Э. Баумана, Москва, 105005, Россия
[3]Университет Соноры, Сонора, 83000, Мексика

*Впервые обсуждаются проблемы оценки индекса подобия неоднородных научных публикаций, содержащих уравнения и формулы. Показано, что наличие уравнений и формул (а также графиков, рисунков и таблиц) является осложняющим фактором, существенно затрудняющим исследование таких текстов. Доказано, что метод определения индекса подобия публикаций, основанный на учете отдельных математических символов и частей уравнений и формул, является неэффективным и может приводить к ошибочным и даже совершенно абсурдным выводам. Исследуются возможности наиболее популярных программных систем Антиплагиат и iThenticate, используемых в настоящее время в научных журналах для выявления плагиата и самоплагиата. Приведены результаты обработки системой iThenticate конкретных примеров и специальных тестовых задач, содержащих уравнения и формулы. Установлено, что эта программная система при анализе неоднородных текстов часто неспособна отличить самоплагиат от псевдосамоплагиата — кажущегося (ложного, мнимого) самоплагиата. Рассмотрена модельная сложная ситуация, в которой идентификация самоплагиата требует привлечения высококвалифицированных специалистов узкого профиля. Предлагаются различные пути улучшения работы программных систем для сопоставления неоднородных текстов. Данная статья будет полезна научным работникам и преподавателям вузов физико-математического и инженерного профиля, программистам, занимающимся проблемой распознавания образов и вопросами цифровой обработки изображений, а также широкому кругу читателей, которые интересуются вопросами плагиата и самоплагиата.*

**1. Введение. Некоторые определения и замечания.** В последнее десятилетие эффективность научных журналов и результаты публикационной активности научных работников и преподавателей вузов часто стали оценивать с помощью различных наукометрических показателей. Появились такие понятия как импакт-фактор и квартиль (для журналов), индекс цитирования, индекс Хирша и процентиль (для авторов публикаций). Все эти показатели весьма субъективны и условны и обладают рядом существенных недостатков [1, 2], а их численные показатели сильно меняются в зависимости от используемой базы данных научного цитирования (Web of Science, Scopus, РИНЦ и др.). В настоящее время к этим показателям добавился еще индекс подобия научных публикаций, который предназначен для выявления плагиата и самоплагиата [3–7].

В данной статье пойдет речь о научных статьях, книгах и других публикациях в различных областях математики и естественных наук, которые содержат значительное количество уравнений и формул. Наличие уравнений и формул является фактором, существенно осложнящим оценку объема заимствований и индекса подобия в таких текстах.

Важно отметить, что оценка индекса подобия и выявление самоплагиата публикаций с уравнениями и формулами является очень важной, непростой и весьма деликатной темой, которая практически не освещена в литературе и не подвергалась широкому обсуждению научной общественности.

Приведем определения некоторых терминов, которые часто используются далее (часть из них вводятся впервые).

*Уравнение* (в математике и физике) – аналитическая запись, состоящая из букв (обычно латинского и греческого алфавитов), математических символов и цифр, которая содержит знак равенства и связывает известные и неизвестные (искомые) величины. Решить уравнение – значит найти входящие в него неизвестные величины, выразив их тем или иным образом через известные величины.

*Формула* – символьно-аналитическая запись, выражающая связь между различными переменными или/и постоянными величинами[1].

*Однородный текст* – текст, состоящий только из отдельных слов и предложений (без уравнений, формул, графиков, рисунков и таблиц).

*Неоднородный текст* – текст, состоящий из отдельных слов, предложений, уравнений и формул (может включать также графики, рисунки и/или таблицы).

[1] Термины уравнение и формула близки по значению и часто не различаются авторами (особенно в англоязычной литературе). Однако при формулировке задач, в которых фигурирует неизвестная величина, принято говорить об уравнениях. Далее к уравнениям и формулам мы относим также неравенства, функции и любые другие математические конструкции, которые записываются в символьно-аналитической форме.

*Плагиат* — прямое заимствовании частей текста статей и книг других авторов без необходимых ссылок. Перефразирование значимого фрагмента чужих произведений путем изменения слов (а также математических символов и букв в научных текстах с уравнениями и формулами) и порядка их следования при сохранении логической структуры аргументации также является плагиатом, если отсутствуют ссылки на работу, которая была использована. О характерных особенностях и способах идентификации плагиата см. [3, 4].

*Самоплагиат* – повторное использование автором значительных, идентичных или почти идентичных частей собственных текстов из более ранних произведений без ссылки на оригинальный источник. О причинах широкого распространения самоплагиата и способах борьбы с ним см. [6, 8].

*Индекс подобия* – число слов в тексте автора, совпадающее с числом слов в источниках, взятых для сравнения, отнесенное к общему числу слов в тексте автора и умноженное на 100%. Это определение справедливо только для однородных текстов, в которых нет уравнений, формул, графиков, рисунков и таблиц.

*Псевдоплагиат* – кажущийся (ложный, мнимый) плагиат. Авторский текст характеризуется оригинальными идеями и/или новым содержанием и значительным числом совпадений отдельных слов, небольших словесных оборотов, отдельных фрагментов уравнений и частей формул, встречающихся в произведениях других авторов. Совпадения носят технический характер (играют второстепенную роль) и не влияют на основное содержание публикации.

*Псевдосамоплагиат* – кажущийся (ложный, мнимый) самоплагиат. Авторский текст характеризуется оригинальными идеями и/или новым содержанием и значительным числом совпадений отдельных слов, небольших словесных оборотов, отдельных фрагментов уравнений и частей формул, встречающихся в других произведениях того же самого автора. Совпадения носят чисто технический характер.

*Антиплагиат* – наиболее известная российская программная система по поиску заимствований в текстовых документах (в научных статьях, диссертациях, дипломных работах, отчетах и др.) из открытых источников в сети Интернет и ряда библиотек [9,10]. Услуги этой системы предоставляются любым зарегистрированным посетителям сайта antiplagiat.ru бесплатно с ограниченной функциональностью или на платной основе с расширенной функциональностью. По поводу корректности и границ использования системы Антиплагиат до сих пор идут обсуждения, споры и судебные процессы [11].

*iThenticate* – это коммерческая программная система обнаружения плагиата и самоплагиата, которая продается издателям, информационным агентствам, корпорациям, юридическим фирмам и пра-вительственным учреждениям, ее клиентами также являются Всемирная организация здравоохранения, Организация Объединенных Наций и Всемирный банк [12]. В настоящее время iThenticate является самой популярной и наиболее мощной системой проверки англоязычных текстов, которая в последнее время активно используется в научных журналах для отсеивания статей с большим индексом подобия[2] на начальном этапе (до рецензирования статей). Краткое описание работы iThenticate и конкретные примеры ее применения можно найти в [14, 15].

В целом использование систем Антиплагиат и iThenticate показало их высокую эффективность для выявления плагиата и самоплагиата научных публикаций, состоящих из однородных текстов, в различных областях социально-гуманитарных и общественных наук, включая экономические и юридические науки.

Отметим, что системы Антиплагиат и iThenticate после обработки статьи выдают ее текст, в котором цветом помечены слова и предложения, совпадающие со словами и предложениями других статей и книг различных авторов, выбранных из соответствующих баз данных, включающих много электронных публикаций. В случае анализа на самоплагиат сравнивается текст рассматриваемой статьи с текстами других публикаций тех же самых авторов.

В математике и естественных науках применение существующих аналитических систем для определения индекса подобия публикаций с неоднородным текстом может привести к ошибочным и даже совершенно абсурдным выводам. Отдельные критические замечания ученых по поводу использования системы iThenticate для статей с уравнениями (формулами) можно найти в научной сети ResearchGate.

Мы с разных сторон протестировали систему iThenticate на ряде математических и физических статей и специальных тестовых задачах, содержащих уравнения, формулы, решения, графики, рисунки и таблицы. Результаты нашего анализа и выводы в систематизированном виде последовательно представлены ниже.

**2. Необходимо исключить из сравнения все, что не связано с научным содержанием статьи.** Имена, фамилии и рабочие адреса авторов, адреса электронной почты, ключевые слова, ссылки на гранты и иную финансовую поддержку, благодарности, фразы о вкладе авторов и об отсутствии конфликта между ними, а также любые другие слова и предложения, не связанные с научным содержанием статьи, не должны учитываться при определении индекса подобия ста-тьи. Несмотря на всю очевидность указанных требований подобная информация в настоящее время часто используется при определении индекса подобия статей с помощью системы iThenticate.

[2] Для англоязычных журналов по математике и физике, индексируемых в базах цитирования Web of Science и Scopus, обычно допустимый индекс подобия статьи не должен превышать 15%. В некоторых журналах допустимый индекс подобия больше и составляет 20–30% [13].

**3. Необходимо исключить из сравнения научные термины и устойчивые словосочетания.** Отдельные научные термины и устойчивые словосочетания, общепринятые в научном сообществе (такие как *элементарная функция, непрерывная функция, скалярное произведение, обыкновенное дифференциальное уравнение, уравнение Лапласа, задача Коши, первая краевая задача, метод коллокаций, метод сращиваемых асимптотических разложений, метод Фурье, теорема существования и единственности, диффузионный пограничный слой, автомодельное решение, число Рейнольдса, евклидово пространство* и многие другие), не относятся к самоплагиату и не должны учитываться при определении индекса подобия статьи, поскольку они не могут быть заменены другими словами без существенного ухудшения текста.

К сожалению, в настоящее время научные термины и устойчивые словосочетания учитываются системой iThenticate, что нередко приводит к существенному завышению индекса подобия статей и незаслуженному обвинению авторов в самоплагиате (научные термины можно убрать при сравнении текстов, но это весьма кропотливая и нудная работа, которую ответственные и технические сотрудники журналов стараются избежать, поскольку ее приходится делать вручную).

Нет никаких разумных оснований приравнивать к самоплагиату также простые и часто используемые в математике краткие словесные обороты и фразы типа: *подставив выражение (1) в уравнение (2), получим; решение задачи (3) имеет вид; рассмотрим нелинейное дифференциальное уравнение; что и требовалось доказать; где $A$ и $B$ – произвольные постоянные; преобразование независимой переменной; на рисунке 1 изображена зависимость; важно отметить, что* и т. д.

Важно отметить, что подавляющее большинство активно публикующихся ученых постепенно вырабатывают свой индивидуальный авторский стиль написания текстов, который заключается в более частом использовании ими конкретных слов, фраз и словесных оборотов, а также выборе способов построения предложений и логических конструкций. При работе над текстами статей эти авторы не переписывают и не копируют отдельные фразы из своих предыдущих публикаций (а именно это им необоснованно ставят в вину разработчики iThenticate), они просто пишут, используя свой стиль. Поэтому при имеющей место в настоящее время практике применения iThenticate обычно наблюдается не выявление самоплагиата, а крайне невежественная и варварская борьба с индивидуальным стилем авторов, что далеко не одно и тоже. Такой подход можно классифицировать как очевидное нарушение элементарных авторских прав (искусственное препятствие возможности индивидуального научного самовыражения и творчества). При этом научная составляющая публикаций отходит на второй план, а основной становится навязываемая авторам малосодержательная оформительская деятельность ненужная для читателей.

Приведем простой аналог, поясняющий нелепость сложившегося положения дел. Каждый бумажный научный журнал имеет свою индивидуальную обложку. Давайте потребуем теперь, чтобы оформление обложки при выходе каждого выпуска (или тома) журнала не менее чем на 85% процентов отличалось от оформления обложек предыдущих выпусков. Такое требование к научным журналам выглядит дико, но полностью соответствует практике работы iThenticate с текстами авторов.

Учитывая вышесказанное и следуя [15], при использовании системы iThenticate надо исключать из проверки короткие последовательности слов (длиною меньше восьми – десяти слов) и списки литературы. Такие опции предусмотрены в iThenticate, однако в журналах крайне редко исключают из рассмотрения короткие последовательности слов из-за неумения работать с этой системой. В [15] отмечается, что исключение коротких последовательностей слов и списка литературы может в полтора – два раза сократить объем выявляемых совпадений даже для простых однородных текстов без формул.

Полезно напомнить, что ученые пишут статьи для своих коллег и заинтересованных специалистов смежных профессий, но никак не для разработчиков системы iThenticate. Читателям важно понять, какие новые результаты получены автором публикации, при этом им глубоко наплевать насколько используемые в статье отдельные слова и фразы по форме отличаются от слов и фраз, написанных тем же автором в его предыдущих статьях. Основная задача авторов – получить новые результаты и написать об этом понятно и четко, а главная цель менеджеров iThenticate – получение максимальных доходов от коллективных и индивидуальных пользователей этой системы (причем интересы бизнеса здесь явно преобладают над здравым смыслом).

**4. Система iThenticate не способна адекватно сравнивать формулы и уравнения.**

***4.1. Качественные особенности и проблема определения индекса подобия статей с уравнениями и формулами.***

1. В однородных публикациях, которые содержат только слова (но не содержат уравнения, формулы, графики, рисунки и таблицы), система iThenticate, как и система Антиплагиат, определяет их индекс подобия в процентах следующим образом[3]. Подсчитывается общее число слов в тексте рассматриваемой статьи, которые совпадают со словами в других статьях, выбранных из привязанных к системе iThenticate баз данных. Затем общее число таких совпадающих слов делится на общее число слов в тексте данной статьи и результат умножается на 100%.

2. В статьях по математике и теоретической физике уравнения и формулы обычно играют основную роль и имеют больший удельный вес, чем сопроводительное словесное описание, которое часто является вторичным и значительно менее важным. Поэтому при исследовании таких публикаций на самоплагиат в первую очередь должны анализироваться уравнения и формулы.

В результате возникает следующий важный вопрос: каким образом можно сопоставить отдельное уравнение или формулу с обычным текстом без формул? Возможны, например, следующие варианты: (i) можно считать формулу эквивалентной слову; (ii) можно считать каждую букву и каждый математический символ, входящие в формулу, эквивалентными одному слову; (iii) можно каждую формулу заменить ее словесным описанием, а затем посчитать число слов в словесном описании. Каждый из указанных вариантов имеет свои недостатки. Вариант (i) неудачен поскольку совершенно обесценивает формулы, часто представляющие собой концентрированную символьно-аналитическую запись, объединяющую различную информацию; вариант (ii) неоднозначен поскольку часто одну и ту же формулу можно записать по-разному (например, $e^x$ и $\exp x$) или представить в виде нескольких формул; последний вариант (iii) неоднозначен (словесное описание может быть разным) и сложен для практической реализации.

3. Наиболее важной характерной качественной особенностью статей с формулами и уравнениями является то, что взаимно однозначная замена всех (или части) букв во всех формулах и уравнениях на любые другие (для этих целей могут использоваться буквы латинского и греческого алфавитов, а иногда и готического алфавита) не меняет содержания статьи. Другими словами, две статьи, которые отличаются только обозначениями букв в формулах и уравнениях, считаются тождественными. Система iThenticate формулы и уравнения, отличающиеся только переобозначениями букв, считает различными[4] (и вряд ли такое положение дел удастся существенно изменить в лучшую сторону в обозримом будущем).

[3] Описывается простейший способ определения индекса подобия (в системе iThenticate предусмотрены некоторые модификации и усложнения подсчета индекса подобия).

Указанное обстоятельство резко ограничивает возможности адекватного применения аналитических систем типа iThenticate для сопоставления отдельных формул в разных текстах. Приходится за основу работы таких систем с неоднородными текстами принять самый простой вариант: две формулы считаются одинаковыми, если все входящие в них буквы и математические символы одинаковы. Далее в разд. 4.2 на конкретных примерах будет показано, что даже при таком простом способе сравнения уравнений и формул в разных текстах iThenticate нередко допускает грубые ошибки, приводящие к существенному завышению процента подобия публикаций.

***4.2. Примеры некорректной работы системы iThenticate.*** Система iThenticate часто не может адекватно сравнивать различные, но внешне похожие формулы и уравнения. Это обстоятельство в настоящее время является одним, но далеко не единственным, из наибольших недостатков этой системы. Ниже приведены несколько иллюстративных примеров того, как система iThenticate ошибочно диагностировала полный плагиат (тождественное совпадение) различных формул и уравнений.

*Пример 1.* Система iThenticate отождествляет две различные формулы

$$g = 1 + |z| + |f|^{1/2} \quad \text{и} \quad g = (1 + |z| + |f|)^{1/2}.$$

(Для доказательства этого см. строку № 3 тестовой задачи 2.)

*Пример 2.* Система iThenticate не различает также формулы

$$y = a + bx^{-1/2} \quad \text{и} \quad y = a + bx - 1/2.$$

(Для доказательства этого см. строку № 4 тестовой задачи 2.)

Из приведенных примеров видно, что iThenticate не умеет работать со скобками и индексами.

*Пример 3.* Система iThenticate показывает, что следующие два разных нелинейных уравнения с частными производными:

$$u_t = [f(u)u_x]_x + g(u) \quad \text{и} \quad u_{tt} = [f(u)u_x]_x + g(u) \qquad (1)$$

являются почти одинаковыми, поскольку у них большинство членов совпадают. Здесь первое уравнение представляет собой нелинейное реакционно-диффузионное уравнение (уравнение параболического типа), а второе уравнение – нелинейное уравнение типа Клейна – Гордона (уравнение гиперболического типа). Мы сравнили две большие статьи [16, 17], содержащие много нумерованных уравне-ний и формул, в которых были построены точные решения более сложных, чем (1), родственных нелинейных уравнений с частными производными, отличающихся только членами $u_t$ и $u_{tt}$ (отметим, что эти уравнения не имеют одинаковых решений). Система iThenticate сделала вывод, что рассматриваемые статьи очень похожи и имеют индекс подобия 61% (расчет основан на подсчете фрагментов неоднородного текста статьи [17], совпадающих с фрагментами текста статьи [16]). Это нелепое заключение, в первую очередь, обусловлено тем, что iThenticate сравнивает отдельные части разных уравнений и формул и считает их частично идентичными, если хотя бы один член или одна буква в них одинаковы (затем все такие псевдосовпадения учитываются при вычислении индекса подобия). Очевидно, что делать общие выводы путем сравнения отдельных составляющих частей формул и уравнений совершенно абсурдно.

[4] Поэтому некоторые авторы для уменьшения индекса подобия статей иногда используют переобозначения букв в формулах и уравнениях.

Для иллюстрации сказанного ниже приведен небольшой фрагмент статьи [17], обработанный системой iThenticate (красным помечены слова и составные части формул, совпадающие со словами и частями формул в статье [16]). Видно, что подавляющее большинство букв и математических символов, определяющих уравнения и формулы, закрашена, т. е. iThenticate учитывает их при вычислении индекса подобия. Для дилетанта эта раскрашенная картинка произведет сильное впечатление и будет служить убедительным доказательством наличия самоплагиата, однако квалифицированный специалист посмотрит сравниваемые статьи и увидит, что все это совершенно несущественно, поскольку уравнение гиперболического типа (75), у которого не закрашена левая часть, сопоставляется с уравнением параболического типа с другой функцией $g(u)$ (т. е. проведенное системой сопоставление было бессмысленным). Закрашенные краткие словесные обороты и стандартный термин не влияют на существо дела (см. разд. 3).

**Solution 12.** By setting $c(x) = 1$, $A = (k+2)/k$, and $B = 1$ in the first three equations of (69), we find that

$$a(x) = x^n, \quad b(x) = x^{n-2}, \quad \eta(x) = \ln x, \quad n = 2 + \frac{2}{k}. \tag{74}$$

This leads to the Klein–Gordon type equation

$$u_{tt} = [x^n f(u) u_x]_x + x^{n-2} g(u), \tag{75}$$

where $n \neq 2$ and

$$g(u) = -\frac{2}{(n-2)h(u)}\left(1 + \frac{2}{n-2}\frac{h'_u(u)}{h^2(u)}\right)\exp\left[-(n-2)\int h(u)\,du\right] - (n-1)\frac{f(u)}{h(u)} - \frac{1}{h(u)}\frac{d}{du}\left[\frac{f(u)}{h(u)}\right], \tag{76}$$

which has the exact solution in implicit form

$$\int h(u)\,du = \ln x + \frac{2}{n-2}\ln t. \tag{77}$$

*Тестовая задача 1.* Чтобы подробнее и лучше продемонстрировать несостоятельность процедуры определения индекса подобия неоднородных текстов с помощью системы iThenticate, рассмотрим теперь комбинированную тестовую задачу, содержащую два варианта неоднородного англоязычного[5] текста, в которых многие слова одинаковы, а рассматриваемые уравнения и полученные решения совершенно разные (см. ниже). Как и ранее, красным цветом выделены совпадающие фрагменты сравниваемых текстов, обнаруженные системой iThenticate.

*Тестовая задача 1, вариант 1.*

Let us consider the pantograph-type parabolic equation with logarithmic nonlinearity

$$u_t = au_{xx} + bu\ln\bar{u} + cu, \tag{i}$$

where $\bar{u} = u(px, qt)$ and $0 < p, q < 1$. We will now prove that equation (i) admits a multiplicative separable solution of the form

$$u(x,t) = f(x)g(t). \tag{ii}$$

Indeed, substituting expression (ii) into equation (i) and separating the variables, we get nonlinear pantograph-type ODEs describing the functions $f = f(x)$ and $g = g(t)$:

$$af''_{xx} + bf\ln\bar{f} + kf = 0,$$
$$g'_t - bg\ln\bar{g} + (k-c)g = 0,$$

where $k$ is an arbitrary constant, and $\bar{f} = f(px)$, $\bar{g} = g(qt)$.

*Тестовая задача 1, вариант 2.*

Let us consider the hyperbolic equation with logarithmic nonlinearity

$$u_{tt} = au_{xx} + bu\ln u + cu. \tag{i}$$

We will now prove that equation (i) admits a multiplicative separable solution of the form

$$u(x,t) = f(x)g(t). \tag{ii}$$

Indeed, substituting expression (ii) into equation (i) and separating the variables, we get nonlinear ODEs describing the functions $f = f(x)$ and $g = g(t)$:

$$af''_{xx} + bf\ln f + kf = 0,$$
$$g''_{tt} - bg\ln g + (k-c)g = 0,$$

where $k$ is an arbitrary constant.

[5] Напомним, что наиболее распространенная версия iThenticate, которую журналы используют для анализа текстов с формулами, работает на английском языке.

Система iThenticate, сравнив текст и уравнения двух вариантов тестовой задачи 1, показывает следующие индексы подобия:

69% – если расчет основан на подсчете фрагментов текста 1-го варианта, совпадающих с фрагментами текста 2-го варианта);

93% – если расчет основан на подсчете фрагментов текста 2-го варианта, совпадающих с фрагментами текста 1-го варианта).

В результате неквалифицированный пользователь системы iThenticate сделает ошибочный вывод: рассмотренные тексты различаются очень слабо.

*Наши комментарии.* Исходные уравнения (i), которые сравниваются, принадлежат к разным типам и совершенно различны: уравнение из варианта 1 является нелинейным функционально-дифференциальным уравнением в частных производных параболического типа с произвольным пропорциональным запаздыванием по двум независимым переменным типа пантографа (впервые примеры точных решений таких уравнений были получены только в 2021 году [19]), а уравнение из варианта 2 является нелинейным уравнением в частных производных гиперболического типа. Функции $f$ и $g$, определяющие точные решения этих уравнений с мультипликативным разделением переменных, также удовлетворяют уравнениям совершенно различного типа: в варианте 1 это нелинейные обыкновенные функционально-дифференциальные уравнения типа пантографа второго и первого порядков, а в варианте 2 – нелинейные обыкновенные дифференциальные уравнения второго порядка. Совпадающие формулы (ii) вообще должны быть исключены из сопоставления, поскольку они являются определением решения с мультипликативным разделением переменных. Короче говоря, содержание этих текстов совершенно отличается друг от друга.

Рассмотренный пример наглядно демонстрирует нелепость использования системы iThenticate для определения индекса подобия научных статей с уравнениями и формулами.

Важно отметить, что связующие слова между уравнениями и формулами в двух сравниваемых вариантах тестовой задачи 1 практически не играют роли. Ученый, написавший несколько статей по точным решениям нелинейных дифференциальных уравнений, поймет содержание обоих вариантов этого примера, если английский текст заменить на немецкий, французский или испанский, и даже если все слова выбросить. Сказанное относится и к любым другим научным статьям, содержащим много уравнений и формул: квалифицированный специалист по теме публикации обычно понимает содержание статьи, даже если из нее выбросить подавляющее большинство слов, но оставить уравнения и формулы (точно также, как многие квалифицированные шахматисты могут играть вслепую без шахматной доски).

*Замечание 1.* При более подробном исследовании результатов анализа тестовой задачи 1 дополнительно обнаружился еще один важный недостаток системы iThenticate. А именно, получаемые индексы подобия существенно зависят от выбора шрифта сопоставляемых неоднородных текстов (выше были приведены индексы подобия для текстов, набранных полужирным курсивом; если использовать обычный курсив, то соответствующие индексы изменятся и будут равны 77% и 94%).

*Тестовая задача 2.* Для дальнейшего анализа возможностей системы iThenticate авторы этой статьи придумали многокомпонентную тестовую задачу, состоящую из двух различных наборов уравнений и формул. В каждом наборе имеется 30 занумерованных уравнений и формул, причем те, которые имеют одинаковый номер (они расположены напротив друг друга) выглядят достаточно похоже по внешнему виду.

Результаты сопоставления указанных двух наборов уравнений и формул представлены ниже. Красным цветом в правом столбце закрашены части формул, которые iThenticate считает идентичными соответствующим частям формул в левом столбце. Итоговый результат, который выдает iThenticate ошеломляет: неоднородный текст в правом столбце на 87% совпадает с неоднородным текстом в левом столбце (расчет основан на подсчете фрагментов текста правого столбца, совпадающих с фрагментами текста левого столбца). Видно, что система iThenticate в пятнадцати случаях вообще не смогла отличить различные уравнения и формулы (что составляет 50% общего тестируемого массива формул). Аналогичным путем было установлено, что неоднородный текст в левом столбце на 71% совпадает с текстом в правом столбце.

Таким образом система iThenticate рекомендует ее пользователю сделать ошибочное заключение о том, что уравнения и формулы в обоих наборах рассматриваемой тестовой задачи очень мало различаются. Поскольку все формулы и уравнения в обоих наборах разные в очередной раз приходим к очевидному выводу: систему iThenticate нельзя использовать для сравнения текстов, которые содержат много уравнений и формул, поскольку эта система неспособна отличить самоплагиат от псевдосамоплагиата (а плагиат от псевдоплагиата).

При сопоставлении неоднородных текстов надо использовать ясный и простой принцип: *два уравнения или формулы считаются одинаковыми, если все входящие в них буквы, математические символы и цифры одинаковы*.

*Тестовая задача 2.*

| | | |
|---|---|---|
| 1 | $y = a + x(b + x)$ | $y = (a + x)(b + x)$ |
| 2 | $y = (a + x)(b + x)$ | $y = (a + x)/(b + x)$ |
| 3 | $g = 1 + \|z\| + \|f\|^{1/2}$ | $g = (1 + \|z\| + \|f\|)^{1/2}$ |
| 4 | $y = a + bx^{-1/2}$ | $y = a + bx - 1/2$ |
| 5 | $y = a + be^{x}$ | $y = a + bex$ |
| 6 | $z = a \sin(x) - 2y$ | $z = a \sin(x - 2y)$ |
| 7 | $y_x = a + bex$ | $yx = a + bex$ |
| 8 | $y' = \exp(ax) - by + c$ | $y' = \exp(ax - by + c)$ |
| 9 | $y' = axy + b(x)y^2$ | $y' = a(x)y + b(x)y^2$ |
| 10 | $y' = a(x)y + b(x)y^2$ | $y' = a(x)y + b(x)y^2 + 1$ |
| 11 | $y' = a(x)y + b(x)y^{k+1}$ | $y' = a(x)y + b(x)y^k + 1$ |
| 12 | $y'' = ay^2 + b$ | $y'' = ay^2 + b + x$ |
| 13 | $y'' = ay^2 + bx$ | $y'' = ay^2 + b(x)$ |
| 14 | $y'' + e^{x}y' + f(y) = 0$ | $y'' + exy' + f(y) = 0$ |
| 15 | $y'' + ay' + by = 0$ | $y'' + ay' + b\|y\| = 0$ |
| 16 | $y'' + axy' + f(y) = 0$ | $y'' + a(xy)' + f(y) = 0$ |
| 17 | $u_t = [f(x)u]_x + g(x)$ | $u_t = [f(x)u_x]_x + g(x)$ |
| 18 | $u_t = f(u)u_{xx} + g(u)$ | $u_t = [f(u)u_x]_x + g(u)$ |
| 19 | $u_t = f(u)u_{xx} + g(u_x)u$ | $u_t = f(u)u_{xx} + g(u_x)/u$ |
| 20 | $u_t = [f(u)u_x]_x + au_x$ | $u_t = [f(u)u_x]_x + aux$ |
| 21 | $u_t = [f(x)u]_{xx} + g(u_x)$ | $u_t = [f(x)u_x]_x + g(u_x)$ |
| 22 | $u_t = [f(x)u]_x + g(u)$ | $u_t = [f(x, u)]_x + g(u)$ |
| 23 | $u_t = [f(x)u]_x + g(u_x)$ | $u_t = [f(x, u)]_x + g(u_x)$ |
| 24 | $u_{tt} = [f(u)u_x]_x + g(u)$ | $u_{tt} = [f(u)u_x]_x + g(u)u_x$ |
| 25 | $u_{tt} = [f(u)u_x]_x + au_x$ | $u_{tt} = [f(u)u_x]_x + aux$ |
| 26 | $u_{tt} = [f(u)u_x]_x + g(u_x)$ | $u_{tt} = [f(u)u_x]_x + xg(u_x)$ |
| 27 | $u_{tt} = [f(u)u_x]_x + g(uu_x)$ | $u_{tt} = [f(u)u_x]_x + g(u, u_x)$ |
| 28 | $u_{tt} = [f(uu_x)]_x + g(u, u_x)$ | $u_{tt} = [f(u, u_x)]_x + g(u, u_x)$ |
| 29 | $u_{xt} = [f(uu_x)]_x + g(u/u_x)$ | $u_{xt} = [f(uu_x)]_x + g(uu_x)$ |
| 30 | $iu_t = au_{xx} + f(\|u\|)u$ | $iu_t = au_{xx} + f(\|u\|)u_x$ |

*Замечание 2.* В тестовой задаче 2, как и в тестовой задаче 1, индексы подобия, получаемые с помощью iThenticate, существенно зависят от выбора шрифта сопоставляемых неоднородных текстов (выше были приведены индексы подобия для уравнений, набранных полужирным курсивом; если использовать обычный курсив, то соответствующие индексы будут равны 78% и 75%).

Во время работы над этой частью статьи одному из авторов приснился страшный сон. А именно, он (профессор) принимает экзамен по математической физике. Студент отвечает плохо и профессор, стремясь натянуть тройку, просит его написать волновое уравнение (т. е. $u_{tt} = a\, u_{xx}$). Студент, немного подумав, пишет уравнение теплопроводности $u_t = a\, u_{xx}$. Естественно профессор ставит ему неуд. Через час профессора вызывают в деканат для обсуждения жалобы студента, которой написал: "Я пропустил в уравнении всего один маленький индекс *t*. Согласно iThenticate индекс подобия написанного мной уравнения составляет целых 87.5%, что является очень хорошим результатом"...

Представьте теперь себе, что абсолютно абсурдную идеологию сравнивания отдельных фрагментов (как это делает iThenticate с уравнениями и формулами) перенесут на художников и их картины. Тогда знаменитый художник И.К. Айвазовский будет объявлен великим плагиатором и самоплагиаторором – действительно, почти на всех его многочисленных картинах изображены вода, волны, тучи и, иногда, корабли. А бедные портретисты: ведь все они рисуют только лоб, щеки, рот, глаза, уши и волосы (иногда и одежду).

Любопытно отметить, что разработчики iThenticate формулы сравнивают по фрагментам, а слова – нет. Спрашивается, почему произошла такая дискриминация формул? Возможно, что отвечающие за эту систему руководящие менеджеры (или ведущие программисты), учась в школе, получали плохие оценки по математике и теперь просто мстят ей за это.

Продемонстрируем, что будет, если слова сравнивать по фрагментам. Возьмем две совершенно разные по смыслу фразы:

"Соврем**енное** дви**жение за** мир" и "Во**енное** поло**жение за**кончилось".

В первой фразе 24 буквы, 12 из них фрагментами (они выделены цветом) совпадают с буквами второй фразы. Поэтому индекс подобия первой фразы составляет 50%.

Вы скажете, что за бред? Но ведь именно так и работает система iThenticate с математическими формулами!

***4.3. Сложные ситуации, требующие привлечения высококвалифицированных специалистов.*** Во многих случаях правильно определить является ли данный коэффициент уравнения несущественным или очень важным может определить только узкий высококвалифицированный специалист по теме рассматриваемой статьи. Более того, адекватные выводы могут быть разными в зависимости от области проводимого исследования. Проиллюстрируем сказанное на простом конкретном примере.

*Пример 4.* Рассмотрим уравнение Абеля второго рода с квадратичной нелинейностью

$$yy'_x - y = ax + bx^2, \qquad (2)$$

где $a$ и $b$ – свободные параметры. Возможны две качественно различные ситуации.

1. Если численно решается задача Коши для уравнения (2), то конкретные значения параметров $a$ и $b$ несущественны. В этом случае два уравнения вида (2) при разных значениях параметров $a$ и $b$ можно считать подобными.

2. Если рассматриваются вопросы интегрируемости уравнения (2), то значения параметра $b$ несущественны, а значения $a$ – весьма существенны. В настоящее время известны всего два значения $a = \pm\frac{6}{25}$, для которых уравнение (2) допускает решение в замкнутой форме [18]. Поэтому если в какой-нибудь статье будет доказана интегрируемость этого уравнения для других значений $a$, этот результат безусловно будет новым. Очевидно, что система iThenticate в данном случае сделает ошибочные выводы.

Этот пример хорошо иллюстрирует принципиальные качественные различия публикаций, посвященным численным и точным решениям математических уравнений. А именно, конкретные значения коэффициентов рассматриваемых уравнений обычно мало существенны при применении численных методов, но, как правило, весьма существенны при построении точных решений.

Дать правильную трактовку результатов обработки статей по интегрируемости и точным решениям обыкновенных дифференциальных уравнений или уравнений с частными производными с помощью системы iThenticate может только высококвалифицированный специалист, специализирующийся в данной области (для анализа корректности или некорректности результатов обработки таких публикаций системой iThenticate нельзя, например, привлекать математиков, которые являются специалистами по численным методам, функциональному анализу и теории чисел). Никакие менеджеры от науки и технические помощники редактора, не имеющие специальных знаний, не способны это сделать.

**5. Система iThenticate не учитывает графики и рисунки.** Система iThenticate не учитывает графики, рисунки и схемы (они просто отбрасываются), что совершенно неправильно. Графики и рисунки часто более важны и наглядны, чем словесный текст статьи, описывающий их. Ошибочность игнорирования рисунков иллюстрируется простым примером.

***Пример 5.*** В книге для начального образования детей имеются два рисунка: на первом изображена кошка, а на втором – собака

(см. ниже). Текст под обоими рисунками один и тот же: «На рисунке изображено домашнее животное, которое имеет четыре лапы, два глаза, два уха, нос, рот, хвост, покрыто шерстью и ест мясо. Напишите в домашней тетради название этого животного». Поскольку система iThenticate не учитывает рисунки, она сделает нелепый вывод, что в этих текстах с разными рисунками имеет место 100% сходство (т. е. кошка = собака). Очевидно, что в рассматриваемом примере рисунки намного важней, чем относящийся к ним текст.

Важно отметить, что иногда графики и рисунки могут составлять основное содержание статьи или быть ее важной неотъемлемой частью[6]. Поэтому их необходимо учитывать при оценке подобия статьи (что технически нетрудно сделать, исходя, например, из занимаемой рисунками площади и площади обрабатываемого iThenticate текста статьи без рисунков).

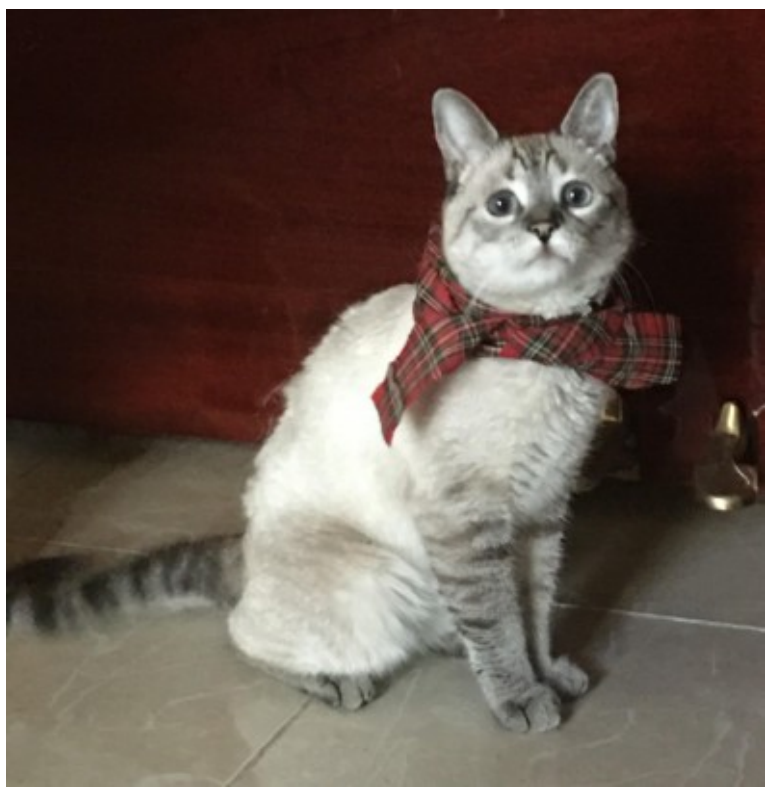

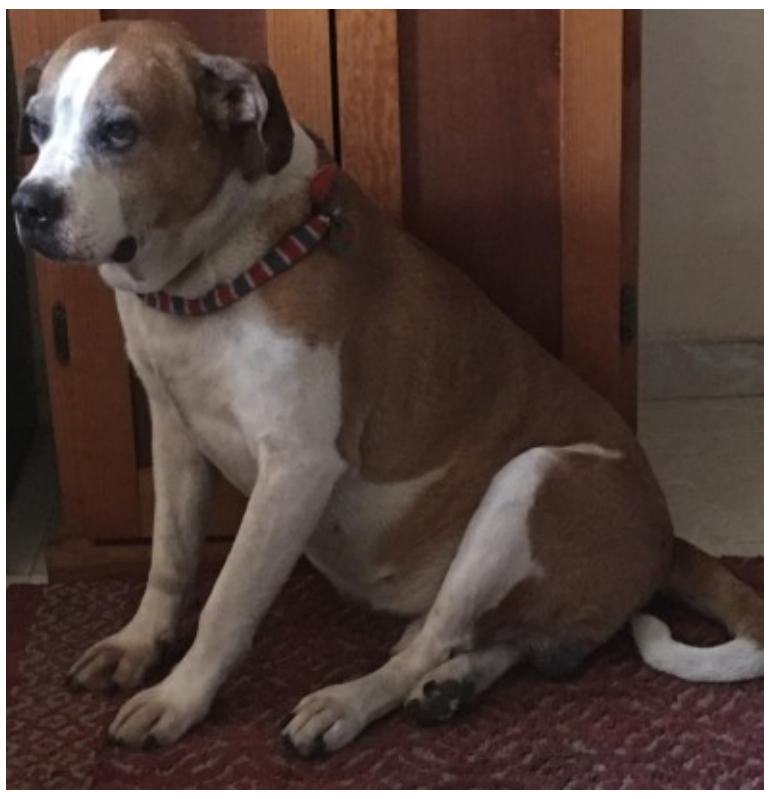

При обработке системой iThenticate математических таблиц, содержащиеся в них уравнения и формулы сравниваются по отдельным фрагментам и буквам, точно также как сравниваются уравнения и формулы в тексте статьи (недостатки такого сравнения подробно описаны в предыдущем разделе).

**6. Возможные пути улучшения работы аналитических систем типа iThenticate с неоднородными текстами, содержащими много уравнений и формул.** Опишем некоторые возможные пути улучшения работы аналитических систем типа iThenticate с неоднородными текстами. Для простоты и наглядности ограничимся рассмотрением неоднородных текстов с большим числом уравнений и формул, но без графиков, рисунков и таблиц.

1. Сначала разработчикам и программистам надо добиться того, чтобы разрабатываемая аналитическая система была способна точно

[6] Особенно это касается экспериментальных работ, а также публикаций, посвященных численному анализу многопараметрических математических моделей и конкретных конструкций и аппаратов.

сравнивать все буквы и математические символы в уравнениях и формулах, содержащихся в разных неоднородных текстах. Для этого надо решить имеющиеся в настоящее время технические проблемы, подробно описанные в разд. 4.2 (см. примеры 1 и 2 и тестовые задачи 1 и 2).

2. Необходимо понять, а затем сформулировать, что и каким образом следует сравнивать в неоднородных научных текстах. Ниже описаны три возможных варианта действий, которые позволят расширить и улучшить возможности существующих и разрабатываемых аналитических систем.

*Вариант 1.* Надо выбросить из рассматриваемой статьи весь словесный текст и сравнивать оставшиеся уравнения и формулы с уравнениями и формулами в других текстах. При этом необходимо использовать указанный в разд. 4.2 принцип сравнения "два уравнения или формулы считаются одинаковыми, если все входящие в них буквы, математические символы и цифры одинаковы".

Этот простой вариант имеет значительное преимущество перед любыми другими, поскольку позволяет сравнивать неоднородные тексты на разных языках (т. к. буквы, специальные математические символы и цифры в уравнениях и формулах не меняются). Важно отметить, что до сих пор такие тексты не поддавались сравнению с помощью существующих аналитических систем.

*Вариант 2.* Надо отдельно сравнить словесный текст и остальной текст, состоящий из уравнений и формул, а затем сложить полученные результаты с подходящими весовыми множителями. При этом весовой множитель для уравнений и формул должен быть выбран значительно большим (в 5–10 раз), чем для словесного текста.

*Вариант 3.* Можно доработать имеющуюся аналитическую систему, учтя замечания, высказанные в разд. 3, и реализовав тот же принцип сравнения, что и в варианте 1.

**7. Краткие выводы.** Приведенные в данной статье логические рассуждения, анализ конкретных примеров и результатов обработки системой iThenticate тестовых задач позволяют сделать вывод о весьма низкой эффективности использования этой системы для оценки индекса подобия неоднородных научных статей, содержащих уравнения, формулы, графики, рисунки и таблицы. В связи с этим необходимо отметить следующее:

1. Нельзя слепо доверять результатам применения системы iThenticate к научным статьям с неоднородным текстом, поскольку цветовые выделения уравнений и формул могут быть ошибочны и их обязательно надо проверять. Проведенное в данной статье обширное исследование дает все основания утверждать, что эта аналитическая система в несколько раз может завышать индекс подобия неоднородных статей.

2. Любые оценки индекса подобия научных статей, содержащих значительное число уравнений, формул, графиков и рисунков, основанные на использовании системы iThenticate (и любых аналогичных аналитических систем, как уже существующих, так и тех, которые появятся в будущем), должны проводиться весьма осторожно высококвалифицированными специалистами по теме рассматриваемых статей.

3. Полные результаты обработки статьи с формулами системой iThenticate (в случае большого индекса подобия) в обязательном порядке должны отсылаться авторам, чтобы у них была возможность проверить эти результаты на адекватность и аргументировано их оспорить.

Другими словами, iThenticate ни в коей мере не решает проблему выявления самоплагиата авторов научных статей с неоднородным текстом, содержащим значительное число уравнений и формул.

Характерной отличительной чертой развития современной науки является то, что часто ученые не могут в полной мере оценить результаты коллег, работающих в казалось бы в очень близких областях (например, специалисты по дифференциальным уравнениям, как правило, не слишком разбираются в интегральных и функциональных уравнениях, более того, специалистов по уравнениям с частными производными обычно нельзя использовать в качестве рецензентов статей по обыкновенным дифференциальным уравнениям, и наоборот). Поэтому квалифицированных рецензентов нельзя заменить административно-техническими работниками и менеджерами, даже если их вооружить аналитической системой сравнения текстов типа iThenticate.

Важно отметить, что недобросовестное выборочное применение существующих аналитических систем к текстам статей некоторых (но не всех) авторов может служить инструментом дискриминации авторов по половому, расовому, национальному и другим признакам.

*Авторы благодарят А.В. Аксенова, А.Л. Левитина и А.Н. Филиппова за внимание к работе и полезные обсуждения.*

ЛИТЕРАТУРА

[1] Игра в цыфирь, или как теперь оценивают труд ученого (сборник статей о библиометрике). М.: МЦНМО, 2011. URL: https://www.mccme.ru/freebooks/bibliometric.pdf

[2] Полянин А.Д. Недостатки индексов цитируемости и Хирша и использование других наукометрических показателей. *Мат. моделирование и численные методы*, 2014, № 1, с. 131–144.

[3] Добрякова Н.И. Заимствование или плагиат: российский и зарубежный

опыт. *Человеческий капитал и профессиональное образование*, 2015, т. 13, № 1, с. 15–20.

[4] Vrbanec T., Mestrovic A. The struggle with academic plagiarism: Approaches based on semantic similarity, 2017 40th International Convention on Information and Communication Technology, Electronics and Microelectronics (MIPRO), 2017, pp. 870–875, doi:10.23919/MIPRO.2017.7973544.

[5] Гельман В.Я. Проблемы формально-механистического подхода к выявлению плагиата в научных работах. *Экономика науки*, 2020, т. 6, № 3, с. 180–185.

[6] Котляров И.Д. Самоплагиат в научных публикациях. *Научная периодика: проблемы и решения*, 2011, № 4, с. 6–12.

[7] Кулешова А.В., Чехович Ю.В., Беленькая О.С. По лезвию бритвы: как самоцитирование не превратить в самоплагиат. Научный редактор и издатель, 2019, т. 4, № 1–2, с. 45–51.

[8] The ethics of self-plagiarism. iThenticate. URL: https://www.ithenticate.com/hs-fs/hub/92785/file-5414624-pdf/media/ith-selfplagiarism-whitepaper.pdf

[9] Антиплагиат. Wikipedia. URL: https://ru.wikipedia.org/wiki/Антиплагиат

[10] Антиплагиат. URL: https://www.antiplagiat.ru/about

[11] Во что превратилась система «Антиплагиат». *The Epoch Times* 19.01.2016; URL: https://www.epochtimes.ru/vo-chto-prevratilas-sistema-antiplagiat-99080032/

[12] iThenticate. Wikipedia. URL: https://en.wikipedia.org/wiki/IThenticate

[13] К сведению авторов. *Сибирский медицинский журнал*, 2018, т. 33, № 4, с. 164.

[14] Grecea M. Similarity (Cross) Check, *Elsevier*. URL: https://www.elsevier.com/__data/assets/pdf_file/0006/865131/Similarity-Cross-Check-webcast-Mihail-Grecea-includes-updated-slid.._.pdf

[15] Рашби Н. От редактора: Вызов Диссернета. *Образование и саморазвитие*, 2017, т. 12, № 1, с. 14–22. URL: https://eandsdjournal.org/wp-stuff/uploads/sites/2/2017/12/122-June-2017.pdf

[16] Polyanin A.D. Construction of exact solutions in implicit form for PDEs: New functional separable solutions of non-linear reaction-diffusion equations with variable coefficients. *Int. J. Non-Linear Mechanics*, 2019, vol. 111, pp. 95–105, doi: 10.1016/j.ijnonlinmec.2019.02.005.

[17] Polyanin A.D. Construction of functional separable solutions in implicit form for non-linear Klein–Gordon type equations with variable coefficients. *Int. J. Non-Linear Mechanics*, 2019, vol. 114, pp. 29–40, doi: 10.1016/j.ijnonlinmec.2019.04.005.

[18] Polyanin A.D., Zaitsev V.F. Handbook of Ordinary Differential Equations: Exact Solutions, Methods, and Problems. Boca Raton – London: CRC Press, 2018.

[19] Polyanin A.D., Sorokin V.G. Nonlinear pantograph-type diffusion PDEs: Exact solutions and the principle of analogy. Mathematics, 2021, vol. 9, no. 5, 511, doi: 10.3390/math9050511.

# The similarity index of mathematical and other scientific publications with equations and formulas and the problem of self-plagiarism identification

A.D. Polyanin[1,2], I.K. Shingareva[3]

[1]Ishlinsky Institute for Problems in Mechanics RAS, Moscow, 119526, Russia
[2]Bauman Moscow State Technical University, Moscow, 105005, Russia
[3]University of Sonora, Sonora, 83000, Mexico

*The problems of estimating the similarity index of inhomogeneous scientific publications containing equations and formulas are discussed for the first time. It is shown that the presence of equations and formulas (as well as figures, drawings, and tables) is a complicating factor that significantly complicates the study of such texts. It has been proved that the method for determining the similarity index of publications, based on taking into account individual mathematical symbols and parts of equations and formulas, is ineffective and can lead to erroneous and even completely absurd conclusions. Possibilities of the most popular software systems Antiplagiat and iThenticate, currently used in scientific journals, are investigated for detecting plagiarism and self-plagiarism. The results of processing by the iThenticate system of specific examples and specific test problems containing equations and formulas are presented. It has been established that this software system, when analyzing heterogeneous texts, is often unable to distinguish self-plagiarism from pseudo-self-plagiarism, seeming real (but false and imaginary) self-plagiarism. A model complex situation is considered, in which the identification of self-plagiarism requires the involvement of highly qualified specialists of a narrow profile. Various ways to improve the work of software systems for comparing inhomogeneous texts are proposed. This article will be useful to researchers and university teachers in physics, mathematics, and engineering, programmers dealing with problems in image recognition and research topics of digital image processing, as well as a wide range of readers who are interested in issues of plagiarism and selfplagiarism.*